\documentstyle[preprint,epsf]{jpsj}

\def\bm#1{{\mbox{\boldmath $#1$}}}

\def\rms#1{{\rm #1}}

\def\bms#1{{\mbox{\rm\scriptsize\it\bf#1}}}

\def\a{\alpha}
\def\b{\beta}

\def\e{\varepsilon}
\def\z{\zeta}

\def\s{\sigma}

\def\bra{\langle}
\def\ket{\rangle}

\def\ua{\uparrow}
\def\da{\downarrow}

\def\runtitle{Hybridization Gap in Kondo Insulators}
\def\runauthor{Tetsuro {\sc Saso} and Hisatomo {\sc Harima}}

\title{%
Formation Mechanism of Hybridization Gap in Kondo Insulators based on a Realistic Band Model and Application to YbB$_{12}$
}

\author{%
Tetsuro {\sc Saso}$^1$ and Hisatomo {\sc Harima}$^2$
}

\inst{%
$^1$Department of Physics, Faculty of Sciences, Saitama University, Shimo-Ohkubo 255, Saitama-City, Saitama 338-8570 Japan.\\
$^2$
The Institute of Scientific and Industrial Research, Osaka University, Mihogaoka 8-1, Ibaraki, Osaka, 567-0047, Japan 
}

\recdate{\hspace{5cm}}

\abst{%
A new LDA+U band calculation is performed on the Kondo insulator material YbB$_{12}$ and an energy gap of about 0.001Ryd is obtained.  Based on this, a simple tight-binding model with 5d$\varepsilon$ and 4f $\Gamma_8$ orbitals on Yb atoms and the nearest neighbor $\sigma$-bonds between them is constructed with a good agreement to the above the LDA+U calculation near the gap.  The density of states is also calculated and the shape is found to be very asymmetric with respect to the gap.  A formation mechanism of the gap is clarified for the first time in a realistic situation with the orbital degeneracies in both conduction bands and the f states.  This model can be a useful starting point for incorporating the strong correlation effect, and for understanding all the thermal, thermoelectric, transport and magnetic properties of YbB$_{12}$.
}

\kword{%
hybridization gap, Kondo insulator, mixing, tight-binding model, LDA+U
}

\begin{document}
\sloppy
\maketitle

\section{Introduction}
Among the rare-earth compounds which include the strongly correlated f electrons, SmB$_6$,\cite{SmB6} YbB$_{12}$,\cite{YbB12} Ce$_3$Bi$_4$Pt$_3$,\cite{Ce3Bi4Pt3} CeRhSb,\cite{CeRhSb} CeFe$_4$P$_{12}$,\cite{CeFe4P12} CeNiSn,\cite{CeNiSn} TmSe\cite{TmSe} etc. exhibit insulating behavior at low temperatures whereas the Kondo-like behaviors (the enhanced electric specific heat and the enhanced paramagnetic susceptibility, etc.) are often observed at higher temperatures.  These materials are called Kondo insulators, Kondo semiconductors or heavy fermion semiconductors.\cite{Riseborough00}  
Recent studies on CeNiSn uncovered that this material is a semimetal with a pseudogap.\cite{CeNiSn99}  TmSe orders antiferromagnetically below $T_\rms{N}$=5K, so that the insulating behavior seems to be due to the gap caused by this order.
Sm has a complicated f shell with five to six f electrons (although the ground state of f$^6$ has vanishing total angular momentum, so that the treatment of Sm ions might not be so difficult than expected\cite{Hanzawa98}).  We exclude these three materials from the following discussions.

The Kondo effect usually occurs in metals, so that the existence of an energy gap at the Fermi energy in insulators may seem contradictory to the occurence of the Kondo effect in these materials.  However, it is already clarified that the Kondo effect can occur if the Kondo temperature $T_\rms{K}$ exceeds the gap size $E_\rms{g}$.\cite{Ogura93}  Even in the compounds with the periodic array of rare-earth ions, the effect of the strong correlation is mainly to renormalize the gap size if the band calculation does yield an energy gap.\cite{Saso96}  Thus, we can understand that the Kondo insulator is a band insulator with a strong correlation.\cite{Aeppli92,Kasaya92}
In fact, all the known Kondo insulators have even number of electrons which fill the bands below the gap.  (TmSe has odd number of electrons, so that it may not be classified into the Kondo insulators also in the present sense.)
In this context, the concept of the Kondo insulator may be extended beyond the rare-earth compounds.
FeSi\cite{Jaccarino67} is considered to be such an example among the transition metal compounds with 3d electrons\cite{Urasaki99} although the correlation may not be so strong as in rare-earth compounds.

Recently, some of the Kondo insulator compounds are attracting renewed interests because of a possibility to be a potential candidate for an efficient thermoelectric device.\cite{Mahan98}  However, quantitative analysis on these materials have been hindered because of the lack of a simple description of the basic electronic structures.  LDA band calculations on the Kondo insulators including the f-electrons as itinerant ones are carried out, giving rising to an energy gap around the Fermi energy,\cite{Takegahara93} or at least a tendency towards the opening of the gap.\cite{Yanase92}  Nevertheless, the obtained band structures look rather complicated, so that a simpler tight-binding model description is necessary to explore the effect of the strong correlation starting from the band structure calculations.

The simplest theoretical model to describe the Kondo insulators is the periodic Anderson model (PAM).  Up to now, mostly the case with only the spin degeneracy has been investigated.\cite{Saso96,Dorin92,Mutou94}  In this case, the conduction band dispersion $\varepsilon_\bms{k}$ and the f level depicted in Fig.1(a) mix up to yield the bands shown in Fig.1(b) after the hybridization, and the energy gap opens.  However, it has already been criticized that this too simple scenario does not work in realistic systems with orbital degeneracy.  It was pointed out by Anderson\cite{Anderson81} that only one orbital among f states can mix with the plane wave state, and other f states remain unmixed.  This situation was treated later in more detail.\cite{Kontani97,Saso01}  Next, consider the case shown in Fig.2(a) where the conduction band has the two-fold orbital degeneracy. (The spin degeneracy is not explicitly shown here and henceforth.)  It is clear from Fig.2(b) that the bands after the hybridization can not have a gap.\cite{Ohkawa84}

Sometimes, a PAM with orbital degeneracy in both conduction bands and f states is considered.\cite{Riseborough92,Ono94}  In this case, however, an artificial assumption that the local symmetry of the f state is conserved even when an electron propagates from site to site, is made in order to facilitate the calculation.

Theoretical studies have been performed also on the model in which the conduction band is treated by the free-electron model but the anisotropic hybridization matrix elements with the f-electron states under the crystalline electric field (CEF) are taken into account.\cite{Ikeda96,Ohara97,Moreno00}  Applicability of this type of model may, however, be limited to the materials which have such a free-electron-like conduction band.  Thus, we need a more sophisticated model and an explanation of the formation of a gap in real materials based on reliable band calculations.

\begin{figure}[bthp]
\epsfxsize=7cm
\centerline{\epsfbox{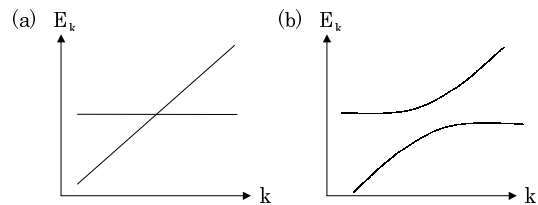}}
\caption{The simplest picture for the formation of the hybridization gap.  The dispersion curves (a) without mixing and (b) after the mixing is introduced.}
\label{fig:Gap1}
\end{figure}
\begin{figure}[bthp]
\epsfxsize=7cm
\centerline{\epsfbox{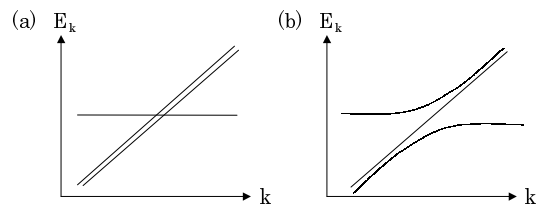}}
\caption{The case that the hybridization gap does not open because of the degeneracy in the conduction band.  The dispersion curves (a) without mixing and (b) after the mixing is introduced.}
\label{fig:Gap2}
\end{figure}

YbB$_{12}$ is the best material to study the formation mechanism of the gap because it is one of the most intensively investigated materials.  Thermal,\cite{Iga} transport,\cite{Sugiyama88} optical,\cite{Okamura} PES,\cite{Susaki} neutron scattering\cite{neutron} experiments have been performed.  In addition, the crystal structure is the simple NaCl type, where Yb ions and B$_{12}$ clusters are located at the interpenetrating fcc sites.  Previous LDA band calculation,\cite{Yanase92} however, resulted in a semimetal with small overlap of the conduction and valence bands.  
Role of the spin-orbit interaction was found to be important by the comparison with the calculation for LuB$_{12}$ without it.\cite{Harima85}
We will report a new calculation using the LDA+U method\cite{Anisimov97} in the next section and find that a gap does open.  To clarify the mechanism of the opening of the gap when the orbital degeneracy exists, we will construct a simple tight-binding model for the conduction band and introduce a hybridization with the 4f $\Gamma_8$ state in \S3 and the formation of the hybridization gap is discussed.  \S4 is devoted to classifying and clarifying the formation mechanism of the gap. Conclusions will be given in \S5.

\section{LDA+U Calculation for YbB$_{12}$}
We have performed a new FLAPW band calculation for YbB$_{12}$ using the LDA+U method\cite{Anisimov97} by adding the term ${\cal H}'= \sum_{imm'\s\s'} |im\s\ket V_{mm'}^{\s\s'}\bra im'\s'|$ to the LDA Hamiltonian, where $V_{mm'}^{\s\s'}=U(\frac{1}{2}-n_{mm'}^{\s\s'})$, $U$ is a parameter and is chosen to be 0.5 Ryd, and  $n_{mm'}^{\s\s'}=\sum_{n{\bf k}} \bra im\s|n\bm{k}\ket\bra n\bm{k}|im'\s'\ket$ is the density matrix between  the orbitals $m,m'$ and the spins $\s,\s'$.  Here, $|n\bm{k}\ket$ is the Bloch function, and we consider only $\ell=3$ for the localized orbital $|im\s\ket$ at site $i$.  Note that the original LDA+U method is extended to the case with the spin-orbit interaction by introducing the off-diagonal matrix elements with respect to the spin indices in $n_{mm'}^{\s\s'}$,\cite{Harima01} which is calculated self-consistently.  

Introduction of the potential $V_{mm'}^{\s\s'}$ pulls down the occupied 4f levels compared to the LDA positions.  In the case of YbB$_{12}$, the resultant 4f level lies below the 5d conduction band.  We think that it is an artifact of the LDA scheme, in which the exchange energy may be overcounted for the 4f electons of heavy Yb atom compared to the 2p electrons of the light B atom, because the 4f electrons are located around the core-electrons which contribute the LDA potential.  This phenomenon can not be corrected even by introducting the LDA+U method. Therefore, we have pulled up the 4f levels additionally by 0.3 Ryd.  Whole the calculation was made self-consistent including this additional shift, and then we obtained the 4f levels located in the conduction band, which has $t_{2g}$ character and consists mainly of 5d$\epsilon$ states on Yb.  We consider that these positions of the 4f levels are reasonable since otherwise we can not obtain the gap.  The 4f levels are split into $\Gamma_6$, $\Gamma_7$ and $\Gamma_8$ under the cubic symmetry with the energies $E_{\Gamma_6} < E_{\Gamma_7} < E_{\Gamma_8}$.  

\begin{figure}[bthp]
\vspace{1cm}
\epsfxsize=7cm
\centerline{\epsfbox{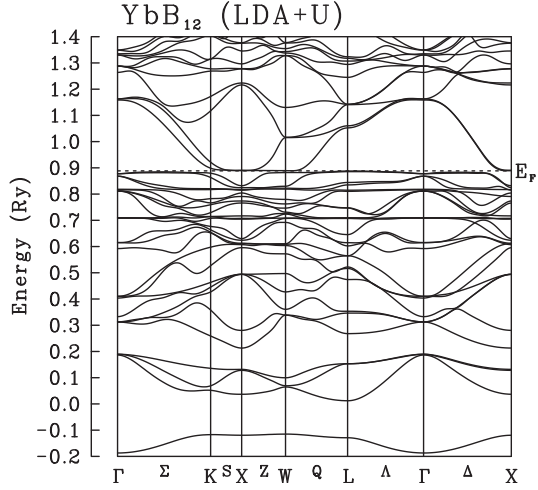}}
\caption{LDA+U band calculation for YbB$_{12}$.  The three flat bands correspond to the 4f $\Gamma_8$, $\Gamma_7$ and $\Gamma_6$ states, respectively.}
\label{fig:BandCalc}
\end{figure}

\begin{figure}[bthp]
\vspace{1cm}
\epsfxsize=7cm
\centerline{\epsfbox{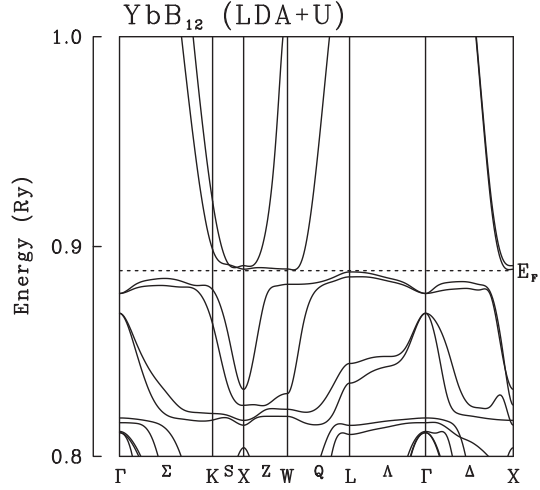}}
\caption{Expanded figure of the dispersion curves near the energy gap.}
\label{fig:BandCalc2}
\end{figure}

\begin{figure}[bthp]
\epsfxsize=7cm
\centerline{\epsfbox{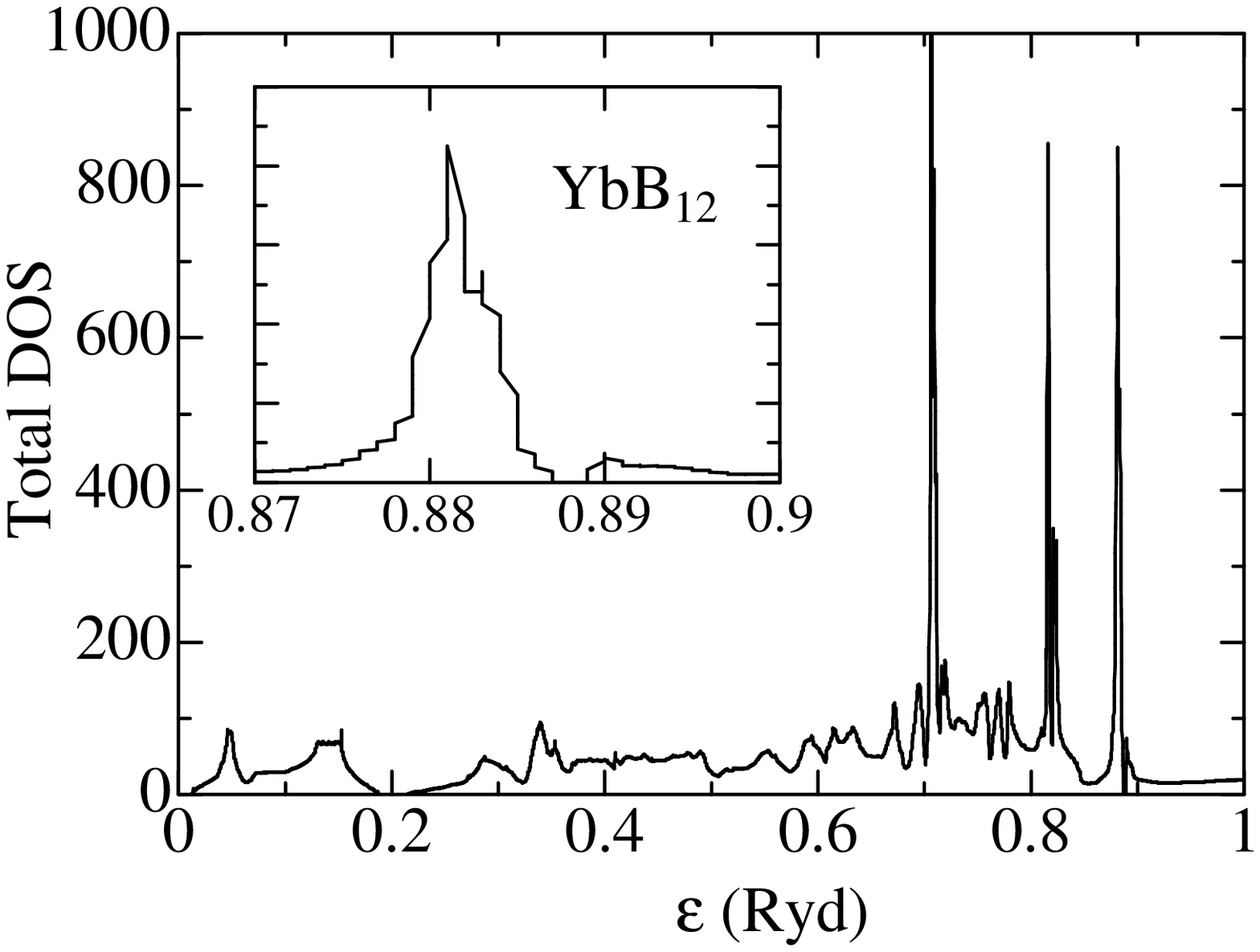}}
\caption{The density of states of YbB$_{12}$ calculated by the LDA+U method.  The insert shows the expanded view of the bands near the gap.}
\label{fig:BandDOS}
\end{figure}

We found that a small gap of about 0.0013 Ryd opens due to the mixing of the 4f $\Gamma_8$ state and the conduction band of the $t_{2g}$ character as shown in Figs.\ref{fig:BandCalc} and \ref{fig:BandCalc2}.  Number of f electrons is about 13.3.  Note that $\Gamma_7$ and $\Gamma_6$ states are lower in energy than $\Gamma_8$, whereas the recent neutron scattering experiment\cite{Alekseev01} found the opposite crystalline field level scheme $E_{\Gamma_6} > E_{\Gamma_7} > E_{\Gamma_8}$.  This contradiction can be simply understood by taking the hole picture.  In this view point, the ground state of 4f$^{13}$ corresponds to the one hole state in $\Gamma_8$, and the excitation to $\Gamma_7$ or $\Gamma_6$ states needs to annihilate this hole and create a hole in $\Gamma_7$ or $\Gamma_6$ states, which needs positive energy in total.  The density of states is displayed in Fig.\ref{fig:BandDOS}. Note that the shape is very asymmetric with respect to the gap.  Whole of the electronic band structure seems very complicated, but if one looks at the dispersion curves near the gap, one finds that it is not so much complicated.  We will express them by a simple tight-binding model in the next section.


\section{Tight-binding Model for YbB$_{12}$}
\begin{figure}[bthp]
\vspace{0.5cm}
\epsfxsize=7cm
\centerline{\epsfbox{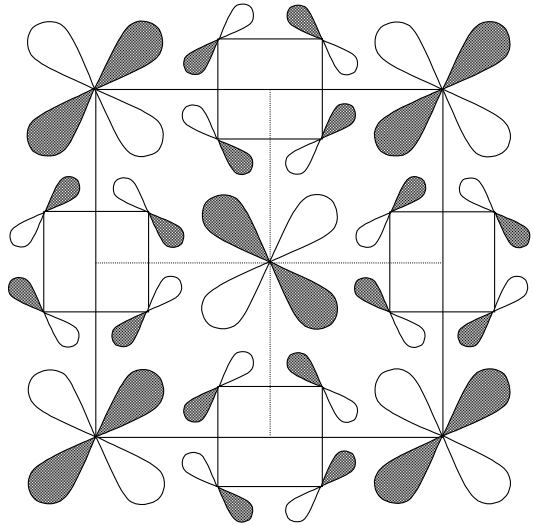}}
\caption{Configuration of the d$\varepsilon$ orbitals (large) on Yb sites and p orbitals (small) on B sites in (001) plane.  The shadows indicate the negative-value parts of the wavefunctions.}
\label{fig:Crystal}
\end{figure}

\begin{figure}[bthp]
\epsfxsize=7cm
\centerline{\epsfbox{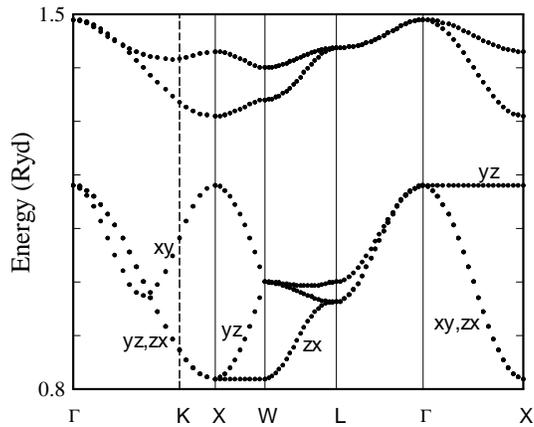}}
\caption{Energy dispersion of the tight-binding model for the YbB$_{12}$ conduction band.  The primary orbital xy, yz or zx in each band is designated for the lower d$\varepsilon$ bands.}
\label{fig:d-band}
\end{figure}

We express the LDA+U bands near the gap by the simple tight-binding model\cite{Slater54} with only the (dd$\sigma$) overlapping integral between 5d$\varepsilon$ ($xy$, $yz$ and $zx$) and 5d$\gamma$ ($x^2-y^2$ and $3z^2-r^2$) orbitals on Yb ions.  This (dd$\sigma$) should be regarded as being produced by the effective hopping through the B$_{12}$ clusters.  Configuration of the orbitals in (001) plane of the crystal is displayed in Fig.\ref{fig:Crystal}.  We locate the energy levels of 5d$\varepsilon$ and 5d$\gamma$ orbitals at $E_{\rm{d}\varepsilon}$=1.0 Ryd and $E_{\rm{d}\gamma}=$1.4 Ryd, respectively, and set (dd$\sigma$)= 0.06 Ryd.
Usually, (dd$\sigma$) is negative, but it is set positive here since the hopping through B$_{12}$ clusters may change the sign, and only this choice can reproduce the LDA+U band calculation.
The resulting bands, shown in Fig. \ref{fig:d-band} along the symmetry axes $\Gamma$(000)-K($\frac{3}{4}\frac{3}{4}\frac{3}{4}$)-X(110)-W(1$\frac{1}{2}$0)-L($\frac{1}{2}\frac{1}{2}\frac{1}{2}$)-$\Gamma$(000)-X(100), consist of the lower 5d$\varepsilon$ and the higher 5d$\gamma$ bands, and look very similar to the dispersions of the conduction bands of $t_{\rm 2g}$ and $\varepsilon_{\rm{g}}$ characters near $E_{\rm F}$ in the LDA and LDA+U calculations, if the f states are removed.

Furthermore, if we take only the d$\varepsilon$ orbitals, the Hamiltonian matrix becomes diagonal and the energy dispersions are given by the following simple expressions:
\begin{equation}
  E_\bms{k}^{\alpha\beta} = E_{d\varepsilon}+3(\rm{dd}\sigma)\cos(\frac{k_\alpha}{2})\cos(\frac{k_\beta}{2}),
  \label{eq:dxy-band}
\end{equation}
where $(\a,\b)=(x,y)$, $(y,z)$ and $(z,x)$.
The splitting at L point in Fig.\ref{fig:d-band} disappeared in this simplified model.  It is easy to see that the lowest band along $\Gamma$-K-X(110) in Fig.\ref{fig:d-band} is doubly degenerate and given by $E_\bms{k}^{yz}$ and $E_\bms{k}^{zx}$ with $k_z=0$, whereas the next band has no degeneracy and is given by $E_\bms{k}^{xy}$ with $k_x=k_y$ and $k_z=0$.  The lowest bands along $\Gamma$-X(100) are also doubly degenerated.

It is expected that the inclusion of the f states at 0.88 Ryd and the mixing with the conduction bands in Fig.\ref{fig:d-band} will yield the bands with the energy gap as obtained by the LDA+U calculation.  
However, the lowest branch in Fig.\ref{fig:d-band} along $\Gamma$-X line is doubly degenerate, so that a simple mixing band picture cannot yield an energy gap, as was mentioned in \S 1.  One has to take account of the symmetry and degeneracy of the 4f states with spin-orbit interaction ($\Gamma_8$) properly as well as those of the conduction bands ($t_{\rm 2g}$).  The band calculation does include these features and yield the energy gap.  Therefore, our tight-binding band will also yield an energy gap if these features are taken into account.  

\begin{figure}[bthp]
\epsfxsize=7cm
\centerline{\epsfbox{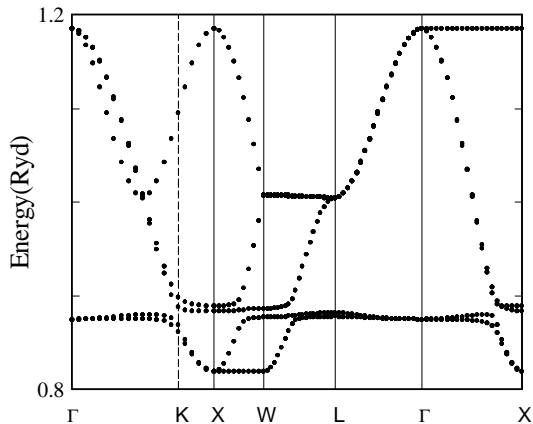}}
\caption{Energy dispersion of the tight-binding model for the YbB$_{12}$ including the hybridization with 4f $\Gamma_8$ states.}
\label{fig:TBBand1}
\end{figure}

\begin{figure}[bthp]
\epsfxsize=7cm
\centerline{\epsfbox{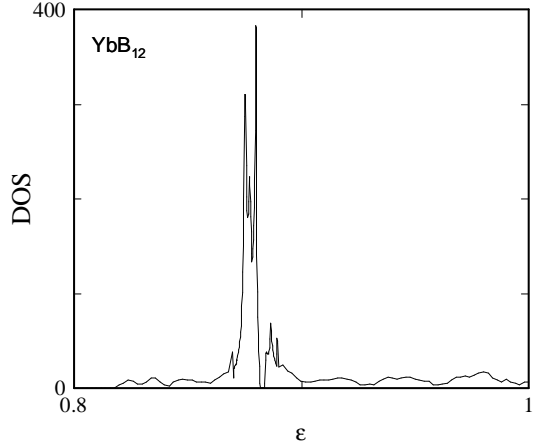}}
\caption{Density of states of the tight-binding model for the YbB$_{12}$ band.}
\label{fig:TBDOS}
\end{figure}

The mixing matrix elements (Slater-Koster integrals) between d and f states have been given only for the d and f states under cubic CEF and without spin-orbit interaction.\cite{Takegahara80}  We need an expression including spin-orbit interaction.
The $\Gamma_8$ states under cubic CEF in the subspace of the total angular momentum $J=7/2$ (for Yb) are expressed in terms of the spherical harmonics $Y_\ell^m$'s with $\ell=3$ and the spinors $\chi_\pm$'s as,
\begin{full}
\begin{subeqnarray}
  |\Gamma_8^{(1)}\pm\ket &=& \left(\sqrt{\frac{7}{12}}Y_3^{\pm 3}-\sqrt{\frac{5}{28}}Y_3^{\mp 1}\right)\chi_\pm -\sqrt{\frac{5}{21}}Y_3^0\chi_\mp, \\
  |\Gamma_8^{(2)}\pm\ket &=& \left(\sqrt{\frac{3}{14}}Y_3^{\pm 2}+\sqrt{\frac{3}{14}}Y_3^{\mp 2}\right)\chi_\pm +\left(\sqrt{\frac{1}{28}}Y_3^{\pm 3}+\sqrt{\frac{15}{28}}Y_3^{\mp 1}\right)\chi_\mp.
\end{subeqnarray}
\end{full}
$Y_3^m$'s can be related to the normalized cubic harmonics by
\begin{full}
\begin{subeqnarray}
  Y_3^0 &=& |z(5z^2-3r^2)\ket, \\
  Y_3^{\pm 1} &=& -\sqrt{\frac{3}{16}}\left[ \mp |x(5x^2-3r^2)\ket -i|y(5y^2-3r^2)\ket \right]
  -\sqrt{\frac{5}{16}}\left[ \pm |x(y^2-z^2)\ket -i|y(z^2-x^2)\ket \right], \\
  Y_3^{\pm 2} &=& \frac{1}{\sqrt{2}}\left[ |z(x^2-y^2)\ket \pm i|xyz\ket \right], \\
  Y_3^{\pm 3} &=& -\sqrt{\frac{5}{16}}\left[ \pm |x(5x^2-3r^2)\ket -i|y(5y^2-3r^2)\ket \right]
  -\sqrt{\frac{3}{16}}\left[ \mp |x(y^2-z^2)\ket -i|y(z^2-x^2)\ket \right]
\end{subeqnarray}
\end{full}
Using these relations, one obtains
\begin{full}
\begin{subeqnarray}
  |\Gamma_8^{(1)}\pm\ket &=& \left[-\frac{1}{2}\sqrt{\frac{5}{21}}\left\{\pm|x(5x^2-3r^2)\ket-i|y(5y^2-3r^2)\ket\right\} \right. \nonumber \\
  & & \left. -\frac{1}{2}\sqrt{\frac{9}{7}}\left\{\mp|x(y^2-z^2)\ket-i|y(z^2-x^2)\ket\right\}  \right] \chi_\pm 
 -\sqrt{\frac{5}{21}}|z(5z^2-3r^2)\ket \chi_\mp, \\
  |\Gamma_8^{(2)}\pm\ket &=& \left[-\frac{1}{2}\sqrt{\frac{5}{7}}\left\{\pm|x(5x^2-3r^2)\ket-i|y(5y^2-3r^2)\ket\right\} \right. \nonumber \\
  & & \left.+\frac{1}{2}\sqrt{\frac{3}{7}}\left\{\mp|x(y^2-z^2)\ket-i|y(z^2-x^2)\ket\right\}  \right] \chi_\mp 
  +\sqrt{\frac{3}{7}}|z(x^2-y^2)\ket \chi_\pm.
\end{subeqnarray}
\end{full}
The mixing matrix elements between the above cubic harmonics and d orbitals under cubic CEF are given in ref.\citen{Takegahara80}.
After lengthy calculations, the matrix elements between d and f states at $\bm{k}$ are finally given by
\begin{full}
\begin{equation}
    \begin{array}{c|cccc}
      & \Gamma_8^{(1)}+ & \Gamma_8^{(1)}- & \Gamma_8^{(2)}+ & \Gamma_8^{(2)}- \\
      \hline
      xy\ua & 5i\sqrt{\frac{5}{56}}t(c_xs_y-is_xc_y) & 0 & 0 & 5i\sqrt{\frac{15}{56}}t(c_xs_y+is_xc_y) \\
      yz\ua & -4\sqrt{\frac{5}{56}}tc_ys_z & i\sqrt{\frac{5}{56}}ts_yc_z & -3i\sqrt{\frac{15}{56}}ts_yc_z & 2\sqrt{\frac{15}{56}}tc_ys_z \\
      zx\ua & -4i\sqrt{\frac{5}{56}}ts_zc_x & i\sqrt{\frac{5}{56}}tc_zs_x & 3i\sqrt{\frac{15}{56}}tc_zs_x & -2i\sqrt{\frac{15}{56}}ts_zc_x \\
      xy\da & 0 & -5i\sqrt{\frac{5}{56}}t(c_xs_y+is_xc_y) & -i\sqrt{\frac{15}{56}}t(c_xs_y-is_xc_y) & 0 \\
      yz\da & i\sqrt{\frac{5}{56}}ts_yc_z & -4\sqrt{\frac{5}{56}}tc_ys_z & 2\sqrt{\frac{15}{56}}tc_ys_z & -3i\sqrt{\frac{15}{56}}ts_yc_z \\
      zx\da & i\sqrt{\frac{5}{56}}tc_zs_x & 4i\sqrt{\frac{5}{56}}ts_zc_x & 2i\sqrt{\frac{15}{56}}ts_zc_x & 3i\sqrt{\frac{15}{56}}tc_zs_x \\
    \end{array}
\label{eq:H}
\end{equation}
\end{full}

\noindent
where $c_\a=\cos(k_a/2)$, $s_\a=\sin(k_\a/2)$ ($\a=x, y, z$) and $t=({\rm df}\sigma$).  Note that we have retained only the nearest neighbor (df$\sigma$) bonds as the simplest model.  Diagonalizing the Hamiltonian matrix for $E_{\Gamma_8}=0.88$ Ryd, (dd$\sigma$)=0.06 Ryd and (df$\sigma$)=0.01 Ryd, we found that a gap almost opens but a small overlap of the bands remains between W and L points.  It was not improved by the introduction of (df$\pi$), (ff$\sigma$) nor the second nearest bonds.  Therefore, we have shifted down the bands below the gap by $\Delta E=-0.005$ Ryd relative to the bands above the gap, similar to the LDA+U treatment, and obtain the dispersion curves shown in Fig.\ref{fig:TBBand1} which have an indirect gap of about 0.003 Ryd between X and L points.  The integrals (df$\pi)=-0.005$ Ryd and (ff$\sigma)=-0.002$ Ryd are also included here to improve the agreement with the band calculation.  Note that (ff$\sigma$) lifts the orbital degeneracy of $\Gamma_8$ states except the symmetry points.  The value of the gap is larger than that in experiment, but we used it for demonstrating the gap clearly.  We also expect that it will be renormalized down to a smaller value by the correlation effect, although the gap size is of course adjustable here.  The density of states is shown in Fig.\ref{fig:TBDOS}, which is very asymmetric with respect to the gap, similar to that obtained by the LDA+U method.


\section{Formation Mechanism of Hybridization Gap}
In order to understand the mechanism of the opening of the gap, we first consider a simple example with degeneracy in both extended and localized orbitals.  $\Gamma_8$ states have four-fold degeneracy, which consists of two-fold orbital degeneracy times two-fold Kramers one.  The Kramers degeneracy always remains even after the hybridization or the inclusion of the spin-orbit interaction.  Therefore, we take $xy$, $yz$ and $zx$ orbitals as the extended states and regard $x^2-y^2$ and $3z^2-r^2$ as the localized ones to imitate the two-fold orbital degeneracy of $\Gamma_8$ states, and construct the tight-binding model with nearest-neighbor (dd$\sigma$) bonds only between d$\varepsilon$-d$\varepsilon$ and d$\varepsilon$-d$\gamma$.  The Hamiltonian matrix reads as
\begin{full}
\begin{equation}
\begin{array}{cc}
  &
  \begin{array}{ccccc}
    \hspace{0cm} xy \hspace{1.8cm} & yz \hspace{1.8cm}& zx \hspace{1cm}& x^2-y^2 \hspace{0.5cm}& 3z^2-r^2
  \end{array} \\
  \begin{array}{c}
    xy \\
    yz \\
    zx \\
    x^2-y^2 \\
    3z^2-r^2
  \end{array}
  &
  \left(
  \begin{array}{ccccc}
     E_\bms{k}^{xy} & 0 & 0 & 0 & \sqrt{3}t_{\varepsilon\gamma}s_xs_y\\
     0 & E_\bms{k}^{yz} & 0 & \frac{3}{2}t_{\varepsilon\gamma}s_ys_z & -\frac{\sqrt{3}}{2}t_{\varepsilon\gamma}s_ys_z \\
     0 & 0 & E_\bms{k}^{zx} & -\frac{3}{2}t_{\varepsilon\gamma}s_zs_x & -\frac{\sqrt{3}}{2}t_{\varepsilon\gamma}s_zs_x \\
     0 & \frac{3}{2}t_{\varepsilon\gamma}s_ys_z & -\frac{3}{2}t_{\varepsilon\gamma}s_zs_x & E_{d\gamma} & 0 \\
     \sqrt{3}t_{\varepsilon\gamma}s_xs_y & -\frac{\sqrt{3}}{2}t_{\varepsilon\gamma}s_ys_z & -\frac{\sqrt{3}}{2}t_{\varepsilon\gamma}s_zs_x & 0 & E_{d\gamma}
  \end{array}
  \right)
\end{array}
\label{eq:dd-band}
\end{equation}
\end{full}

\begin{figure}[bthp]
\epsfxsize=4cm
\centerline{\epsfbox{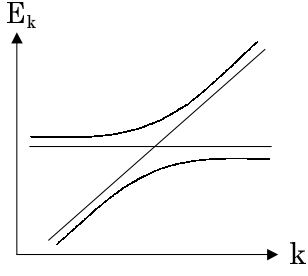}}
\caption{The dispersion curves corresponding to the Hamiltonian matrix eq.(\ref{eq:Gap5}) is shown.}
\label{fig:Gap5}
\end{figure}

\begin{figure}[bthp]
\epsfxsize=7cm
\centerline{\epsfbox{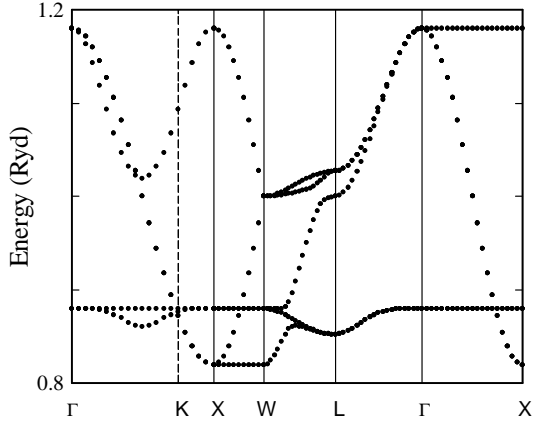}}
\caption{The tight-binding band with d$\varepsilon$-d$\varepsilon$ and d$\varepsilon$-d$\gamma$ integrals only. The parameters are chosen as $t_{\varepsilon\gamma}=0.5(dd\sigma)$.}
\label{fig:TBBand2}
\end{figure}

\begin{figure}[bthp]
\epsfxsize=7cm
\centerline{\epsfbox{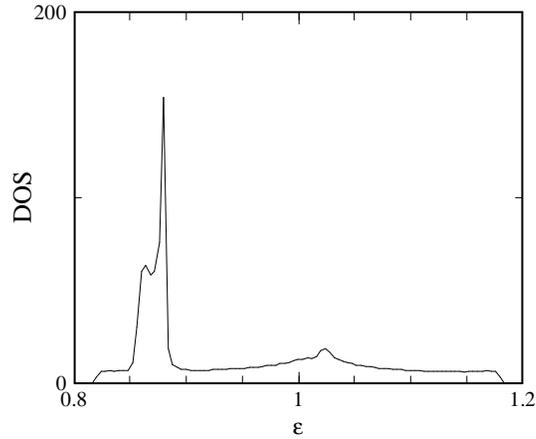}}
\caption{The density of states of the tight-binding band shown in Fig.\ref{fig:TBBand2}.}
\label{fig:TBDOS2}
\end{figure}

\noindent
$E_\bms{k}^{\alpha\beta}$ are given by eq.(\ref{eq:dxy-band}).  Usually, $t_{\varepsilon\gamma}$ must be equal to (dd$\sigma$) but here we assume $t_{\varepsilon\gamma} \neq (dd\sigma)$.  In this model, the mixing matrix elements are proportional to $\sin(k_y/2)\sin(k_z/2)$ and $\sin(k_z/2)\sin(k_x/2)$ for the lowest yz and zx bands, respectively, so that the gap vanishes along the symmetry axes $\Gamma$-K-X(110)-W(1$\frac{1}{2}$0) and $\Gamma$-X(100) where $k_z=0$, as shown in Fig.\ref{fig:TBBand2}.  The density of states is shown in Fig.\ref{fig:TBDOS2} and becomes singular at the flat band.  Note that a gap does open, of course, if we replace $\sin(k_\alpha/2)$ by a constant value in eq.(\ref{eq:dd-band}),  but it does not open if we replace all the off-diagonal matrix elements between d$\gamma$'s and d$\varepsilon$'s by the same value of a constant.  This is because of an accidental degeneracy, and can be more easily seen by considering the simpler Hamiltonian matrix
\begin{equation}
  \left(
  \begin{array}{cccc}
    \e_\bms{k} & 0 & V & V \\
    0 & \e_\bms{k} & V & V \\
    V & V & E_f & 0 \\
    V & V & 0 & E_f \\
  \end{array}
  \right),
  \label{eq:Gap5}
\end{equation}
which has the eigenvalues $\e_\bms{k}$, $E_f$ and $[\e_\bms{k}+E_f\pm\sqrt{(\e_\bms{k}-E_f)^2+16V^2}]/2$. The dispersion curves look like Fig.\ref{fig:Gap5}.

Note also that if the Hamiltonian matrix is given by
\begin{equation}
  \left(
  \begin{array}{cccc}
    \e_\bms{k} & 0 & V & 0 \\
    0 & \e_\bms{k} & 0 & V \\
    V & 0 & E_f & 0 \\
    0 & V & 0 & E_f \\
  \end{array}
  \right),
  \label{eq:Gap6}
\end{equation}
the eigenvalues are the hybridized bands $[\e_\bms{k}+E_f\pm\sqrt{(\e_\bms{k}-E_f)^2+16V^2}]/2$, each of which is doubly degenerated and has a gap.  This or similar models have been often used in the literature,\cite{Riseborough92,Ono94} but are not appropriate since the local symmetry of the f state is conserved even through the hopping from one site to the other.

\begin{figure}[bthp]
\epsfxsize=7cm
\centerline{\epsfbox{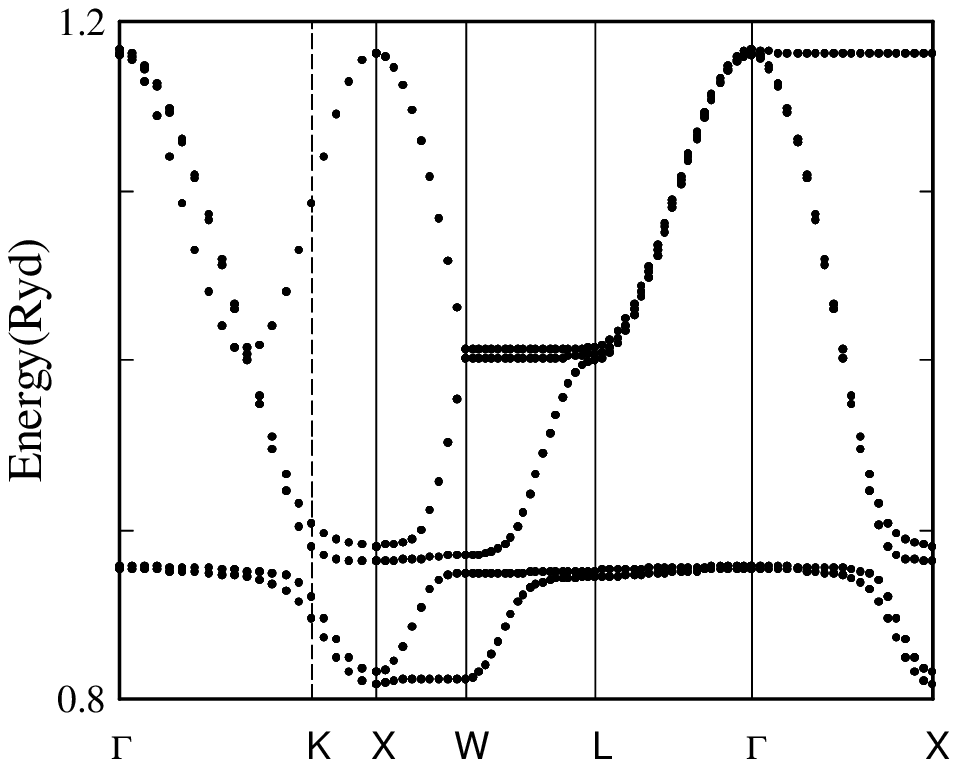}}
\caption{Energy dispersion of the tight-binding model for the YbB$_{12}$ in which all $s_\alpha$'s and $c_\alpha$'s are replaced with unity.  This calculation does not include (df$\pi$) nor (ff$\sigma$).}
\label{fig:TBBand3}
\end{figure}

Based on these analyses, we consider eq.(\ref{eq:H}) again.  The structure of this matrix looks rather similar to that of eq.(\ref{eq:dd-band}), although the inclusion of the spin-orbit interaction makes it complex.  If we set all $s_\alpha=c_\alpha=1$ in eq.(\ref{eq:H}), then we obtain a gap of about 0.003 Ryd without an LDA+U shift $\Delta E$ (Fig.\ref{fig:TBBand3}).  On the other hand, gap does not open because of accidental degeneracy if we give the same value of a constant to all the off-diagonal matrix elements.

\begin{figure}[bthp]
\epsfxsize=7cm
\centerline{\epsfbox{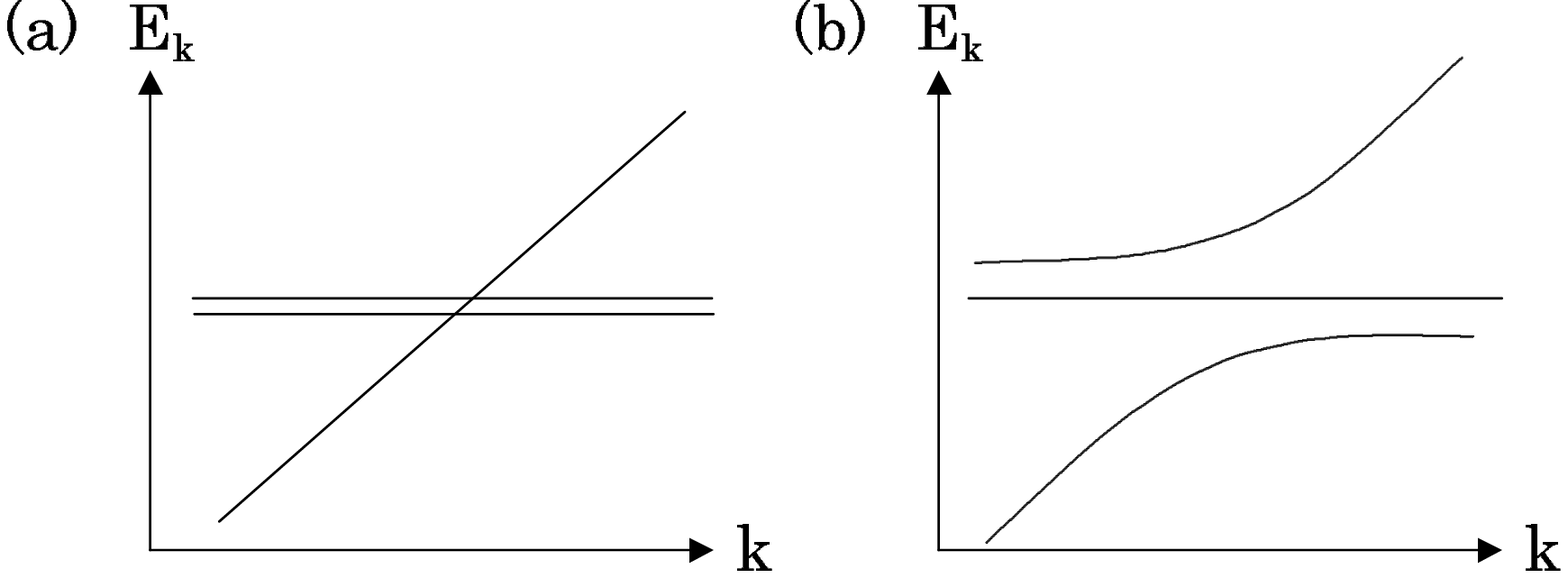}}
\caption{The case that the hybridization gap does open.  The dispersion curves (a) without mixing and (b) after the mixing is introduced.}
\label{fig:Gap3}
\end{figure}

\begin{figure}[bthp]
\epsfxsize=7cm
\centerline{\epsfbox{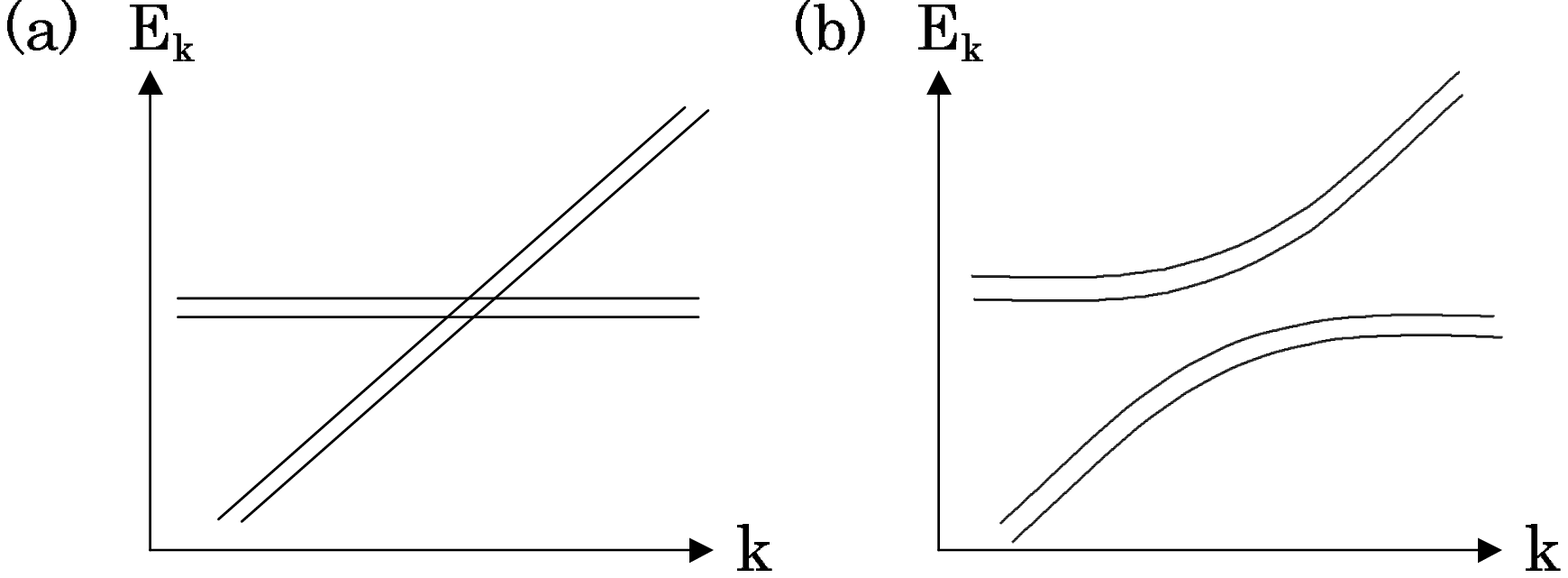}}
\caption{The case that the hybridization gap does open.  The dispersion curves (a) without mixing and (b) after the mixing is introduced.}
\label{fig:Gap4}
\end{figure}

We can summarize the formation mechanism of an energy gap in the Kondo insulators as follows.
As shown in Fig.\ref{fig:Gap3}, if there is a two-fold orbital degeneracy (besides the Kramers one) in the f states and no degeneracy in the crossing conduction band, a gap can open, but there is a possibility that the bands might overlap due to some dispersion of the localized states.  When both the conduction band and the f state have two-fold degeneracy, a gap can also open as shown in Fig.\ref{fig:Gap4} except the case of accidental degeneracy or the strong dispersion in f state.  
Furtheremore, complex off-diagonal matrix elements due to the spin-orbit interaction makes it difficult for the gap to close.

On the other hand, gap does not open if the f state has no orbital degeneracy as in $\Gamma_6$ or $\Gamma_7$ states and the conduction band has degeneracy (Fig.\ref{fig:Gap2}).  Therefore, the experimental fact that there is a gap in YbB$_{12}$ indicates that the ground state is $\Gamma_8$.  (Note if $\Gamma_7$, say, is above the gap, and $\Gamma_8$ is at the gap, the gap opens, but the number of f electrons in the ground state becomes twelve, which can not be accepted based on the experimental facts.)


\section{Conclusions}
  We have performed a new LDA+U band calculation for the most typical Kondo insulator YbB$_{12}$ and obtained a gap of about 0.0013 Ryd.  Based on this calculation, we constructed a simple tight-binding band model to express the bands near the energy gap. The conduction band consists of the 5d$\varepsilon$ orbitals on Yb and the effective overlap integral (dd$\sigma$) is regarded as being produced by the hopping through the B$_{12}$ clusters.  This model can describe the $t_{\rm 2g}$ conduction band very well. Inclusion of the mixing with the 4f $\Gamma_8$ states resulted in the energy gap after additional shift of the filled bands similar to the LDA+U treatment.  The formation mechanism of the gap in the realistic situation with the degeneracy of the conduction bands and the f states is classified and clarified for the first time.  

The present model is very useful in constructing a theory with correlation effect, and may consistently explain all the thermal, thermoelectric, transport and magnetic properties of YbB$_{12}$.
Such a calculation is now in progress.  Preliminary calculation for the thermoelectric power is already reported,\cite{Saso02} in which the asymmetry of the density of states with respect to the gap plays an important role in the temperature dependence of the thermopower.  Finally, it should be emphasized that the f-f hopping is also important, although small, to reproduce the band calculation, so that a use of a too simplified model with the plane wave conduction electrons and the completely localized f-electrons may need take care.  Especially, anomalies in the density of states\cite{Ikeda96,Ohara97,Moreno00} obtained for such a simplified model may be sensitive to the inclusion of the f-f hopping.

%
%
\acknowledgement
The author thanks Dr. H. Kontani and Dr. T. Mutou for useful conversations.

\end{document}